\documentclass[prl, twocolumn,amsmath,amssymb,superscriptaddress]{revtex4}
\usepackage{graphicx, subfigure}
\usepackage{dcolumn}
\usepackage{color}
\usepackage{bm}
\usepackage[T1]{fontenc}
\usepackage{balance}

\newcommand{\bket}[1]{\left<#1\right>}
\newcommand{\scprod}[2]{\left<#1|#2\right>}
\newcommand{\ket}[1]{\left|#1\right>}
\newcommand{\bra}[1]{\left<#1\right|}

\newcommand{\ba}{\text{\bf{a}}}
\newcommand{\bb}{\text{\bf{b}}}
\newcommand{\bH}{\text{\bf{H}}}

\begin{document}

\title{Deterministic protocol for mapping a qubit to coherent state superpositions in a cavity}
\author{Zaki Leghtas}
\affiliation{INRIA Paris-Rocquencourt, Domaine de Voluceau, B.P.~105, 78153 Le Chesnay Cedex, France}
\author{Gerhard Kirchmair}
\affiliation{Department of Physics and Applied Physics, Yale University, New Haven, Connecticut 06520, USA}
\author{Brian Vlastakis}
\affiliation{Department of Physics and Applied Physics, Yale University, New Haven, Connecticut 06520, USA}
\author{Michel Devoret}
\affiliation{Department of Physics and Applied Physics, Yale University, New Haven, Connecticut 06520, USA}
\author{Rob Schoelkopf}
\affiliation{Department of Physics and Applied Physics, Yale University, New Haven, Connecticut 06520, USA}
\author{Mazyar Mirrahimi}
\affiliation{INRIA Paris-Rocquencourt, Domaine de Voluceau, B.P.~105, 78153 Le Chesnay Cedex, France}
\affiliation{Department of Physics and Applied Physics, Yale University, New Haven, Connecticut 06520, USA}

\date{\today}

\begin{abstract}
We introduce a new gate that  transfers an arbitrary state of a qubit into a superposition of two quasi-orthogonal coherent states of a cavity mode, with opposite phases. This qcMAP gate is based on conditional qubit and cavity operations exploiting the energy level dispersive shifts, in the regime where they are much stronger than the cavity and qubit linewidths. The generation of  multi-component superpositions of quasi-orthogonal coherent states, non-local entangled states of two resonators and multi-qubit GHZ states can be efficiently achieved by this gate.
\end{abstract}

\pacs{Valid PACS appear here}
\maketitle


In the field of quantum Josephson circuits, microwave resonators are extremely useful for performing readout, coupling multiple qubits and protecting against decoherence~\cite{Wallraff2005,Majer2007,Paik-et-al-PRL2011}.
In addition, using an oscillator as a memory to store a qubit state has been explored both theoretically and experimentally (see e.g~\cite{gottesman-et-al-01, Mariantoni2011, Maitre1997}). The recent improvement in coherence times of microwave resonators with respect to superconducting qubits~(\cite{Megrant2012,Reagor-al_2012}) makes it particularly interesting to use a cavity as a quantum memory in this context.

In this letter we introduce a new gate between a qubit and a cavity (qcMAP) which maps the qubit state onto a superposition of two quasi-orthogonal coherent states with opposite phases. This gate provides access to the large Hilbert space of the cavity, so that one can encode the information of a multi-qubit system on a single cavity mode and decode it back on the qubits. In particular, this gate can be employed to efficiently prepare any superposition of quasi-orthogonal coherent states (SQOCS)~\cite{raimond-et-al-2010}. Furthermore, we show that this scheme can be easily adapted to prepare entangled states of two resonators, which would maximally violate Bell's inequality.  Finally, the qcMAP gate offers a new method to use the cavity as a {bus} to perform multi-qubit gates and prepare arbitrary GHZ states.

Previously realized qubit-cavity encoding in superconducting qubits are based on a correspondence between the register's states and the cavity's Fock states~\cite{martinis-nature08}. These schemes are based on bringing the qubit into resonance with the cavity. The qcMAP gate does not require such real-time frequency tunings and hence avoids an extra decoherence channel. Additionally, in contrast to resonant regime, the time to prepare a SQOCS using the qcMAP gate does not increase with the amplitudes of the coherent components, but scales only linearly with the number of coherent components. Large SQOCS could hence be generated with high fidelities, exploring the decoherence of highly non-classical states~\cite{brune-et-al-96,deleglise-et-al:nature08}.

We place ourselves in the strong dispersive regime, where both the qubit and resonator transition frequencies split into well-resolved spectral lines indexed by the number of excitations in qubit and resonator~\cite{schuster-nature07}. The resonator frequency $\omega_r$ splits into two well resolved lines $\omega_r^g$ and $\omega_r^e$, corresponding to the cavity's frequency when the qubit is in the ground ($\ket{g}$) or the excited ($\ket{e}$) state. Through the same mechanism, the qubit frequency $\omega_q$ splits into $\{\omega_q^n\}_{n=0,1,2,\cdots}$ corresponding to the qubit frequency when the cavity is in the photon number state $\ket{n}$. Recent experiments have shown dispersive shifts that are about 3 orders of magnitude larger than the qubit and cavity linewidths~\cite{Paik-et-al-PRL2011}.
\begin{figure}[h]
\setlength{\unitlength}{1mm}
\begin{picture}(120,70)
\put(0,68){{(a)}}
\put(0,36){\includegraphics[width=\columnwidth]{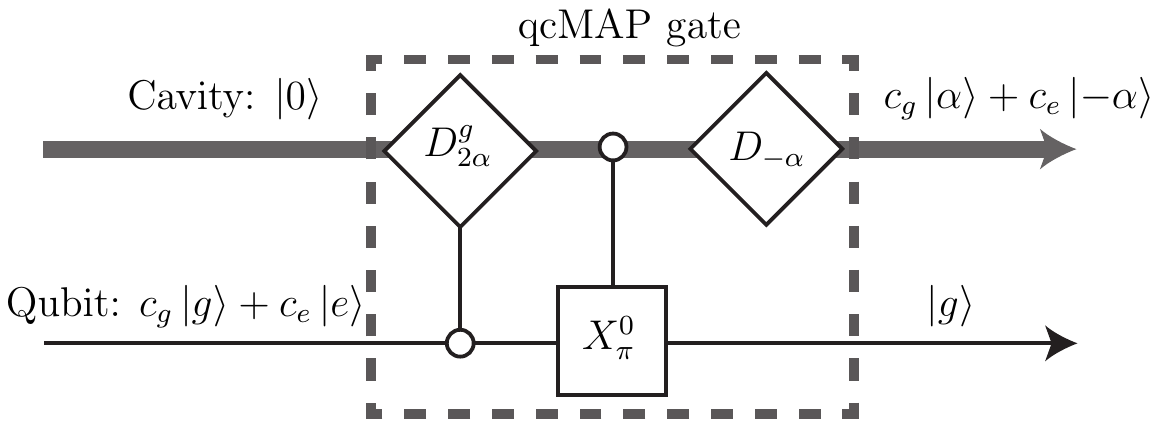}}
\put(0,34){{(b)}}
\put(0,0){\includegraphics[width=\columnwidth]{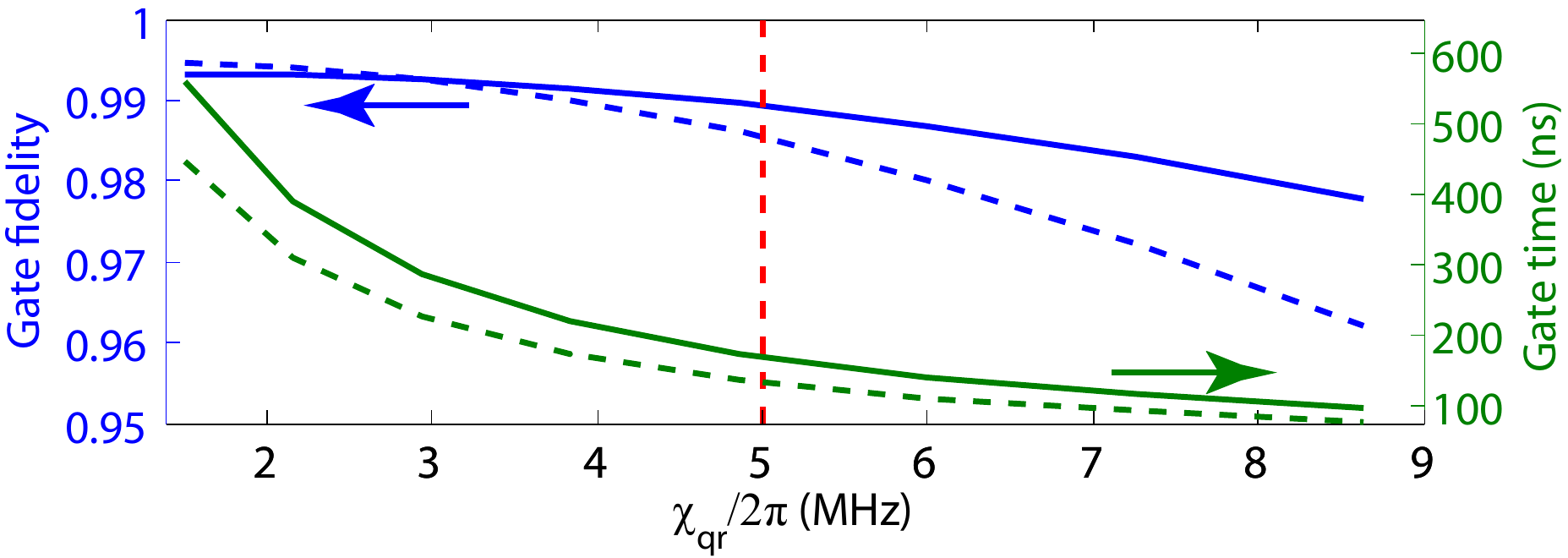}}
\end{picture}
\caption{\label{fig:SwapGate} (a) The qcMAP gate comprises a conditional displacement of the cavity mode $D_{2\alpha}^g$ and a conditional rotation of the qubit $X_\pi^0$, mapping the qubit state to a superposition of two coherent states with opposite phases in the cavity. (b) Fidelity (blue) and gate time (green) of the qcMAP gate as a function of the dispersive coupling $\chi_{qr}$, for two values $3.5$ (solid line) and $7$ (dashed line) of $\bar n=|\alpha|^2$. Increasing $\chi_{qr}$ decreases the gate time, however it also increases the cavity self-Kerr $\chi_{rr}$ which reduces the fidelity. This effect is more important for large coherent states, which explains the more important fidelity drop for $\bar n=7$ photons. For $\bar n=3.5$ photons, fidelities larger than $99\%$ are obtained for $\chi_{qr}$ smaller than $5~$MHz, with a gate time of $\approx 170$~ns, much shorter than achievable coherence times.}
\end{figure}

\begin{figure*}[ht!]
\begin{center}
       \includegraphics[width=.75\columnwidth]{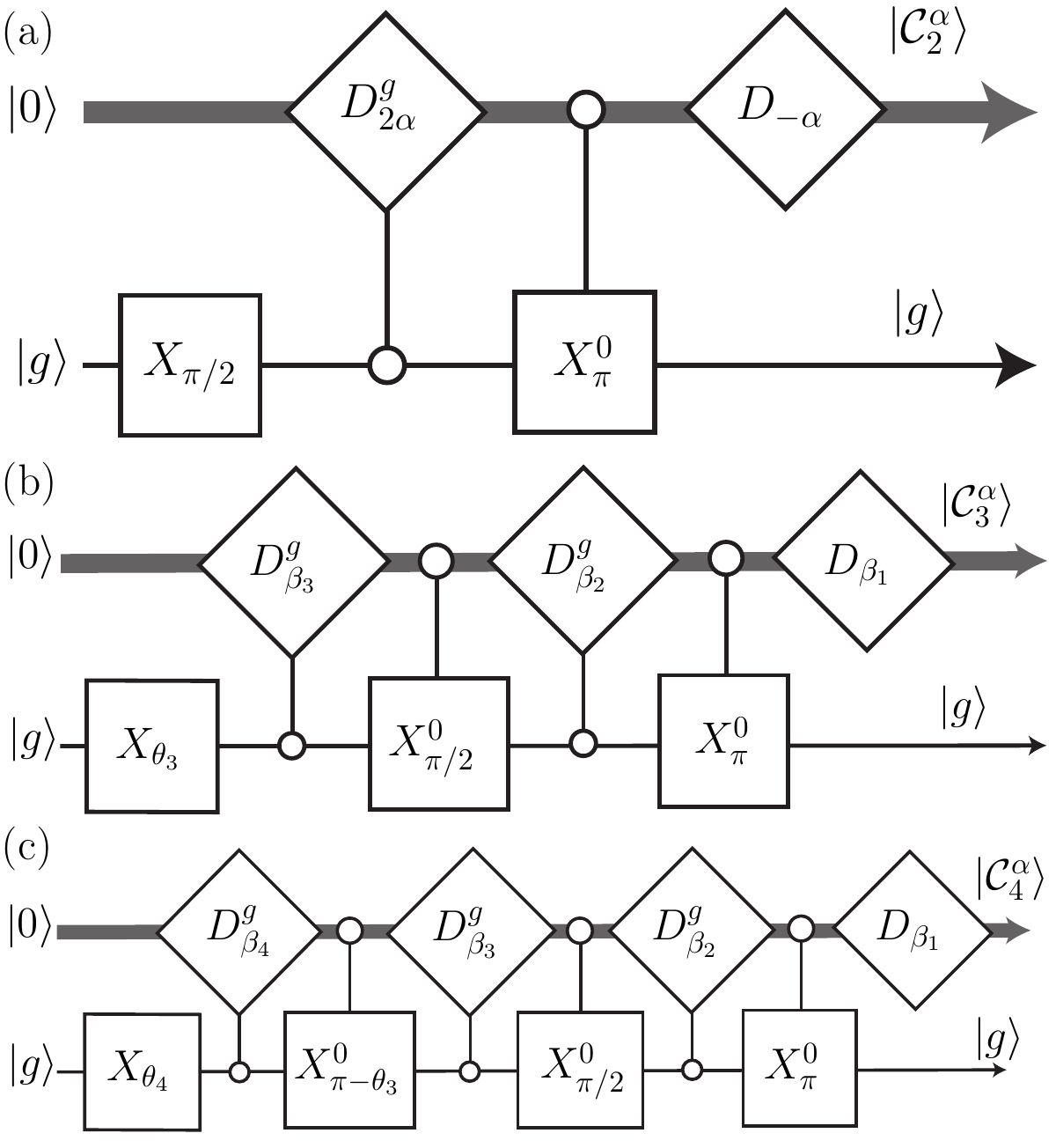}\hspace{-.15cm}
        \includegraphics[width=1.3\columnwidth]{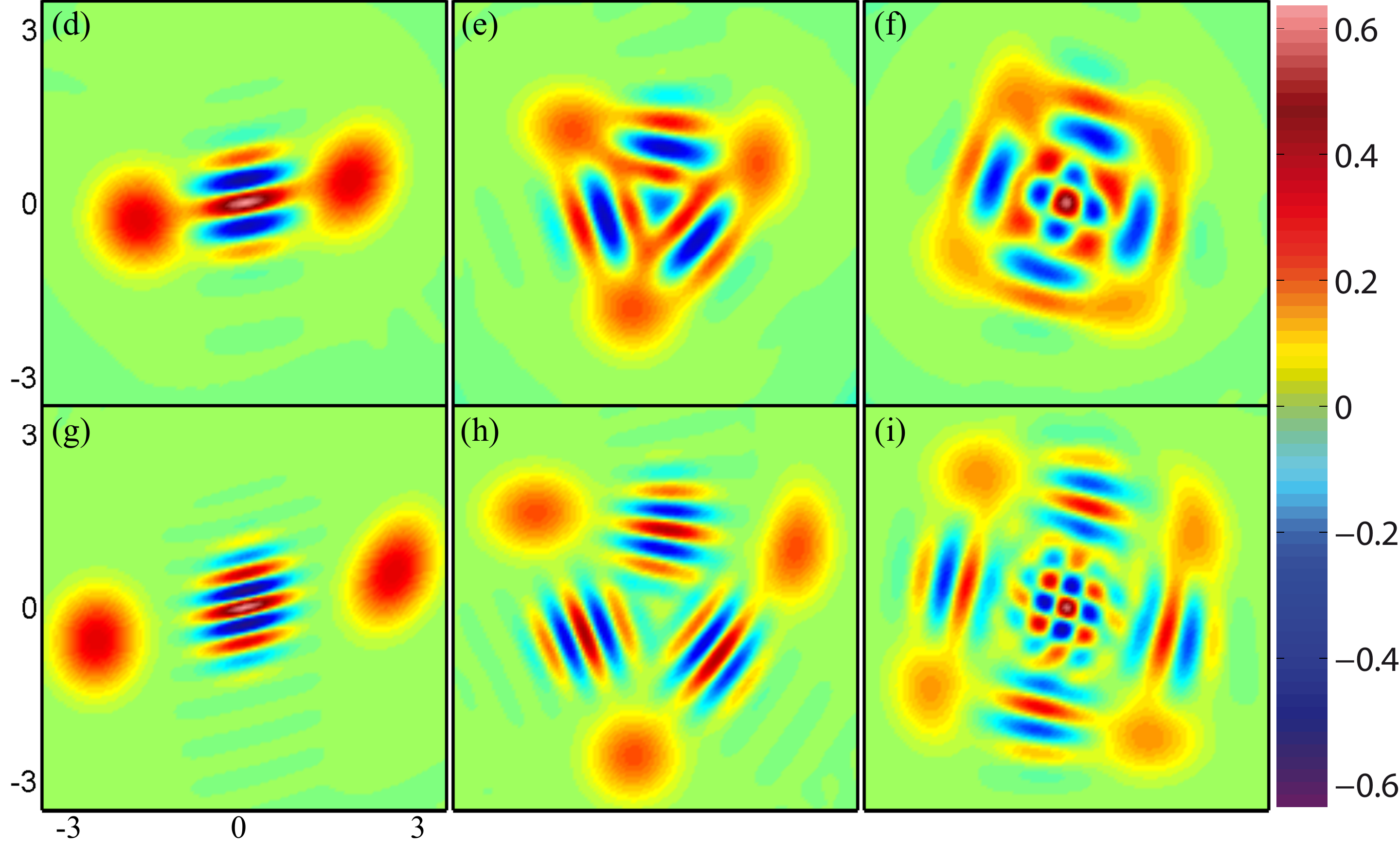}
 \caption{ \label{fig:cats} (a-c) Operations to prepare a 2 (a), 3 (b) and 4 (c) component SQOCS. $\ket{\mathcal{C}^\alpha_n}$ denotes a superposition of coherent states $\ket{\alpha_1}+\ldots+\ket{\alpha_n}$. In (b) and (c),  $\beta_1=\alpha_1$, $\beta_2=\alpha_2-\alpha_1$, $\beta_3=\alpha_3-\alpha_2$ and $\theta_3=2\arccos(1/\sqrt 3)$. In addition, in (c), $\beta_4=\alpha_4-\alpha_3$ and $\theta_4=2\arccos(1/2)$. (d-i) Wigner functions of the prepared states in presence of decoherence and the cavity self-Kerr. The upper figures correspond to $\bar n=3.5$ photons in each coherent component and the lower ones correspond to $7$ photons. We define the fidelity of the prepared state $\ket{\psi_\text{prep}}$ to the target $\ket{C^\alpha_n}$ as $F_{\text{prep}}(\ket{C^\alpha_n})=|\scprod{\psi_\text{prep}}{C^\alpha_n}|^2$. We get $F_{\text{prep}}(\ket{\mathcal{C}^\alpha_2})=97.8\%$ (resp. $97.2\%$) for $\bar n=3.5$ (resp. $\bar n=7$) for a preparation time $T_{\text{prep}}=170$ ns (resp. 135 ns). Similarly, $F_{\text{prep}}(\ket{\mathcal{C}^\alpha_3})=96.2\%$ (resp. $95.7\%$) and $T_\text{prep}(\ket{\mathcal{C}^\alpha_3})=320$~ns (resp. $225$~ns); $F_{\text{prep}}(\ket{\mathcal{C}^\alpha_4})=91.9\%$ (resp. $91.5\%$) and $T_\text{prep}(\ket{\mathcal{C}^\alpha_3})=460$~ns (resp. $355$~ns). Note the insensitivity of the preparation fidelity to the size of the coherent components. Due to the cavity self-Kerr, the components that are created earlier are deformed more than those created later.}
\end{center}
\end{figure*}

The qcMAP gate relies on two operations which we detail in the following: the conditional cavity displacement, which we denote by $D_\alpha^g$, and  the conditional qubit rotation, which we denote by $X_\theta^0$ (see Fig.~\ref{fig:SwapGate}(a)). An \emph{un}conditional displacement $D_\alpha$ is obtained by applying a very short pulse, which displaces a coherent state by $\alpha$ regardless of the qubit state.
A conditional displacement $D_\alpha^g$ can be reaIized in the strong dispersive limit: with a selective pulse of duration $T\gtrsim 1/\chi_{qr}$, we may displace the cavity by a complex amplitude $\alpha$ only if the qubit is in the ground state. For a coherent state $\ket{\beta}$, we have $D_\alpha^g\ket{e,\beta}=\ket{e,\beta}$, and $D_\alpha^g\ket{g,\beta}=e^{(\alpha\beta^\dag-\alpha^\dag\beta)/2}\ket{g,\beta+\alpha}$. Such a conditional displacement was first proposed in~\cite{davidovich-et-al:PRL93} as part of a non-deterministic scheme to prepare a two-component superposition of coherent states. For the deterministic qcMAP gate, we combine this displacement with a conditional qubit rotation $X_{\pi}^0$. The conditional rotations $X_{\theta}^0$ are simply achieved by applying a selective pulse at $\omega_q^0$, performing an rotation of angle $\theta$ of the qubit state conditioned on the cavity being in its vacuum state. Such selective qubit rotations have been experimentally demonstrated in~\cite{johnson-et-al:Nature2010}.

In order to map the state of the qubit to the cavity mode, we construct the qcMAP gate as follows. Starting from a qubit in $c_g\ket{g}+c_e\ket{e}$ and cavity in $\ket{0}$, a first conditional displacement $D_{2\alpha}^g$ entangles the qubit and the cavity, creating the state $c_g\ket{g,2\alpha}+c_e\ket{e,0}$. We choose $2|\alpha|$ to be large enough so that the non-orthogonality of the two coherent states $|\bket{2\alpha~|~0}|^2=e^{-4|\alpha|^2}$ is negligible (of order $10^{-6}$ for $\bar n=|\alpha|^2=3.5$). A conditional $\pi$-pulse $X_\pi^0$ can then disentangle the qubit from the cavity leaving the qubit in $\ket{g}$ and the cavity in $c_g\ket{2\alpha}+c_e\ket{0}$. Finally, the unconditional displacement $D_{-\alpha}$ centers the superposition at the origin.\\

The qcMAP gate is well adapted to quantum information processing with a transmon qubit~\cite{Koch-et-al-07} coupled to a microwave resonator. The Hamiltonian is well approximated by~\cite{nigg-et-al-2012}
\begin{equation*}
\frac{\bH}{\hbar}=\omega_r \ba^\dag \ba+\omega_q \bb^\dag \bb- \frac{\chi_{rr}}{2}(\ba^\dag \ba)^2- \frac{\chi_{qq}}{2}(\bb^\dag \bb)^2 -\chi_{qr}\ba^\dag \ba~\bb^\dag \bb\;.
\end{equation*}
Here $\ba$ and $\bb$ are respectively the dressed mode operators of the resonator and the qubit ($\ket{g}$ and $\ket{e}$ are the first two eigenstates of $\bb^\dag\bb$), $\omega_r$ and $\omega_q$ are their frequencies, $\chi_{qr}$ is the dispersive qubit-resonator coupling, and $\chi_{qq}$ and $\chi_{rr}$ the anharmonicities. Indeed, due to the coupling to a non-linear medium (the qubit), the cavity also inherits a Kerr effect that leads to the anharmonicity $\chi_{rr}=\chi_{qr}^2/4\chi_{qq}$~\cite{nigg-et-al-2012}. This nonlinearity can distort coherent states and sets a limit on the fidelity of the gate.

While an unconditional cavity displacement $D_\alpha$ can be performed rapidly using a short pulse, the conditional cavity displacements $D^g_\alpha$ and qubit rotations $X_\theta^0$ necessitate long pulses allowing to selectively address the corresponding spectral line. In the qcMAP gate, $X_\pi^0$ transforms $\ket{e,0}$ to $\ket{g,0}$ while leaving $\ket{g,2\alpha}$ unchanged. To this end, we apply a pulse with a carrier frequency $\omega_q^0$ and shape it such that it does not overlap with the spectral lines $\omega_q^n(=\omega_q^0-n\chi_{qr})$ corresponding to the qubit frequencies when the cavity is in $\ket{2\alpha}$. Defining $\bar n=\bra{\alpha}\ba^\dag \ba\ket{\alpha}=|\alpha|^2$, the pulse length needs to be longer than a certain multiple of $1/4\bar n\chi_{qr}$. Here we take a Gaussian pulse of standard deviation $\sigma_t=5/4\bar n\chi_{qr}$ and total length $6\sigma_t$ resulting in a $\pi$-pulse time of $15/2\bar n\chi_{qr}$ ($\approx 70$ns for $\bar n=3.5$ and $\chi_{qr}/2\pi=5$MHz) for $99\%$ fidelity. For the $D_\alpha^g$ operation, using a gaussian pulse to selectively address $\omega_r^g$ without driving $\omega_r^e=\omega_r^g-\chi_{qr}$ (the spectral lines are separated by $\chi_{qr}$ and not $4\bar n\chi_{qr}$) would require a relatively long time of $\approx30/\chi_{qr}$. However, as detailed in the supplemental material, $D^g_\alpha$ can be performed using two unconditional displacements and a waiting time between them; the whole operation time is significantly reduced to $\pi/\chi_{qr}$ ($\approx 100$ns). The total gate time is $T_{\text{Gate}}\approx  \frac{15+2\bar n\pi}{2\bar n\chi_{qr}}$ ($T_{\text{Gate}}\approx170$ns).

There is a compromise between decreasing the gate time with larger coupling strengths and increasing the undesirable effect of the cavity self-Kerr. The  Kerr effect leads to a phase collapse of a coherent state with mean photon number $\bar n$ on a time scale of $T_{\text{collapse}}= \frac{\pi}{2\sqrt{\bar n}\chi_{rr}}$~\cite[Section 7.2]{haroche-raimond:book06}. This phase collapse can be considered as an extra dephasing of the cavity and reduces the gate fidelity.

In Fig.~\ref{fig:SwapGate}(b), we compute the fidelity and time of the qcMAP gate in presence of the cavity self-Kerr but without any decoherence. We take $\chi_{qq}/2\pi=300$~MHz and vary $\chi_{qr}$. The fidelity $F$ of the gate $\mathcal U$ is defined as $F=\min_{c_g,c_e}{|(c_g^\dag\bra{g,\alpha}+c_e^\dag\bra{g,-\alpha}) ~\mathcal U~(c_g\ket{g,0}+c_e\ket{e,0})|^2}$. The gate fidelity and the gate time decrease with increasing $\chi_{qr}$. The decrease in fidelity is slightly worse for higher $\bar n$ since the coherent state becomes more exposed to the cavity's non-linearity. The maximum fidelity of $\approx99.5\%$ is set by the fidelity of the conditional $\pi$-pulse which can be arbitrarily improved using longer pulses (at the expense of longer gate times). In presence of decoherence, one should increase the coupling strength (and therefore decrease the gate time) up to values that make the phase collapse due to the cavity self-Kerr comparable to other dephasing times.

One can tailor any SQOCS by applying a sequence of qcMAP gates (see~\cite{raimond-et-al-2010} for another method based on the dynamical quantum Zeno effect, and \cite{Lutterbach-Davidovich-PRA_2000} for a non deterministic scheme). The protocols to generate 2, 3 and 4-component SQOCS are given in Fig.~\ref{fig:cats}(a-c). The master equation simulation of these preparation protocols lead to the Wigner functions shown in Fig.~\ref{fig:cats}(d-i). The corresponding parameters are $\chi_{qr}/2\pi=5$~MHz, $\chi_{qq}/2\pi=300$~MHz and $\chi_{rr}/2\pi=20$~kHz. The qubit relaxation and dephasing times are $T_1=T_2=20~\mu$s, and the cavity decay time is $T_{\text{cav}}=100\mu$s. Recent experiments with transmon qubits  coupled to 3D resonators~\cite{Paik-et-al-PRL2011,Rigetti-et-al-2012} indicate that such parameters are realistic. More details on the preparation scheme can be found in the supplemental material. In particular, one notes the insensitivity of the fidelity to the size of the coherent components. The ability to prepare multi-component SQOCS also implies that the qcMAP gate can be used to store \emph{multi}-qubit states in the resonator.

\begin{figure}
     \begin{center}
\includegraphics[width=.7\columnwidth]{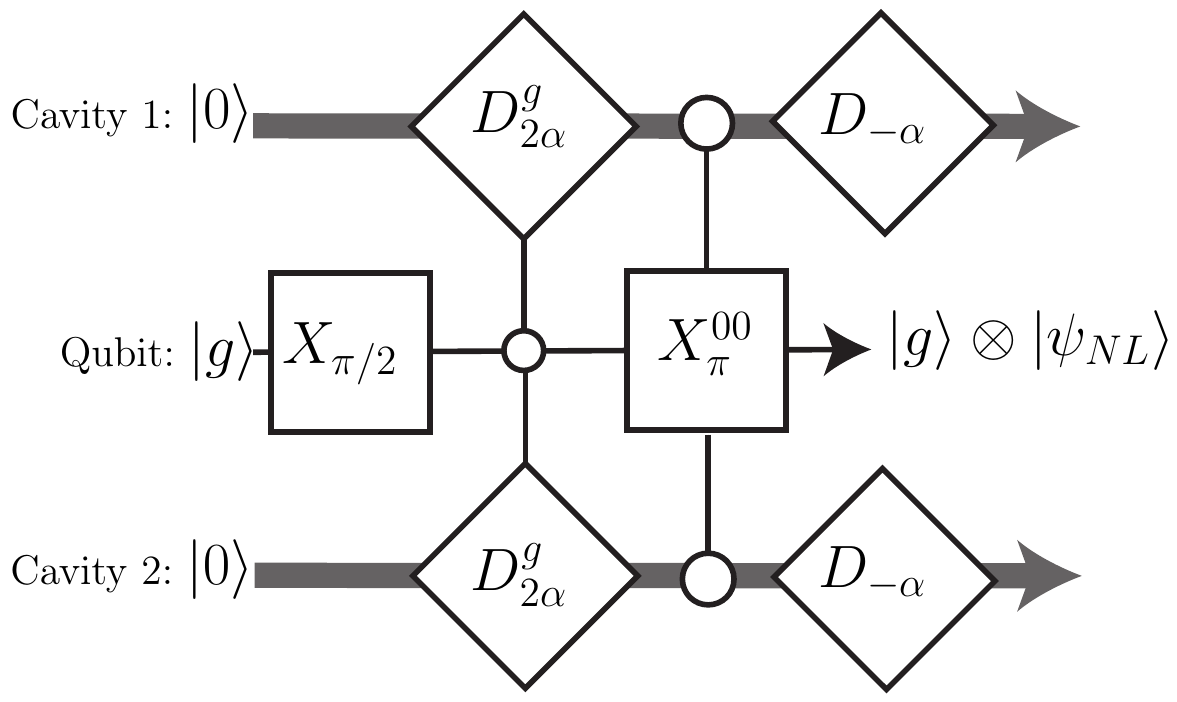}
\caption{\label{fig:NonLocal}  Protocol for preparing a non-local entangled state between two cavities that are dispersively coupled to a single qubit. Two simultaneous conditional displacements lead to a tripartite entanglement, preparing the state $\ket{g,2\alpha,2\alpha}+\ket{e,0,0}$; A $\pi$-pulse on the qubit, conditioned on both cavities being in vacuum, will then disentangle the qubit from the cavities leaving them in an entangled state $\ket{\psi_\text{NL}}=(\ket{-\alpha,-\alpha}+\ket{\alpha,\alpha})/N$, where $N$ is a normalization constant. We obtain a fidelity of $\approx96\%$ in $190$ns, leading to a Bell signal of $2.5$.}
\end{center}
\end{figure}

The qcMAP gate can also be used on a qubit coupled to two spatially separated cavities~\cite{Kirchmair-al_2012} to prepare \emph{non-local} mesoscopic superposition states of the form $\ket{-\alpha,-\alpha}+\ket{\alpha,\alpha}$. Such highly non-classical states achieve a maximum violation of Bell's inequality as soon as $|\alpha|^2\approx 2$~\cite[Section 7.6]{haroche-raimond:book06}. The preparation scheme is sketched in Fig.~\ref{fig:NonLocal}. As in the single-mode case, the sequence duration is set by the length of the selective operations. The two conditional displacements are performed simultaneously and their time is given by $\max(\pi/\chi_{qr_1},\pi/\chi_{qr_2})$ ($\chi_{qr_1}$ and $\chi_{qr_2}$ being the dispersive coupling between the qubit and cavity modes). The conditional $\pi$-pulse is performed in a time of order $15/2\bar n(\chi_{qr_1}+\chi_{qr_2})$. Therefore, the preparation time for a non-local superposition is even shorter than the single-mode case. However, in addition to the cavity self-Kerr effects $\chi_{r_1r_1}$ and $\chi_{r_2r_2}$, we also have a cross-Kerr term $\chi_{r_1r_2}\ba_1^\dag \ba_1~\ba_2^\dag \ba_2$ between the two modes ($\chi_{r_1r_2}$ given by $2\sqrt{\chi_{r_1r_1}\chi_{r_2r_2}}$~\cite{nigg-et-al-2012}).

We simulate this scheme taking $\chi_{qr_1}/2\pi=5$~MHz, $\chi_{qr_2}/2\pi=4$~MHz, $\chi_{r_1r_2}/2\pi=20$~kHz, $\chi_{r_1r_1}/2\pi=20$~kHz, $\chi_{r_2r_2}/2\pi=13$~kHz and $\chi_{qq}/2\pi=300$~MHz, and coherence times of $T_1=T_2=20~\mu$s for the qubit and $T_{\text{cav}}=100\mu$s for the two cavities. The entangled state $\ket{\alpha,\alpha}+\ket{-\alpha,-\alpha}$ with $|\alpha|^2=1.5$  is prepared with a fidelity of $\approx96\%$ in 190 ns. By measuring the two-mode Wigner function at four points, as explained in~\cite{banaszek-wodkiewicz-99,milman-et-al-2005,sarlette-et-al-2012}, we retrieved a Bell signal of 2.5, largely violating Bell's inequality (maximum possible Bell signal is $2\sqrt 2$).

\begin{figure}[h]
\setlength{\unitlength}{1mm}
\begin{picture}(120,70)
\put(0,36){\includegraphics[width=\columnwidth]{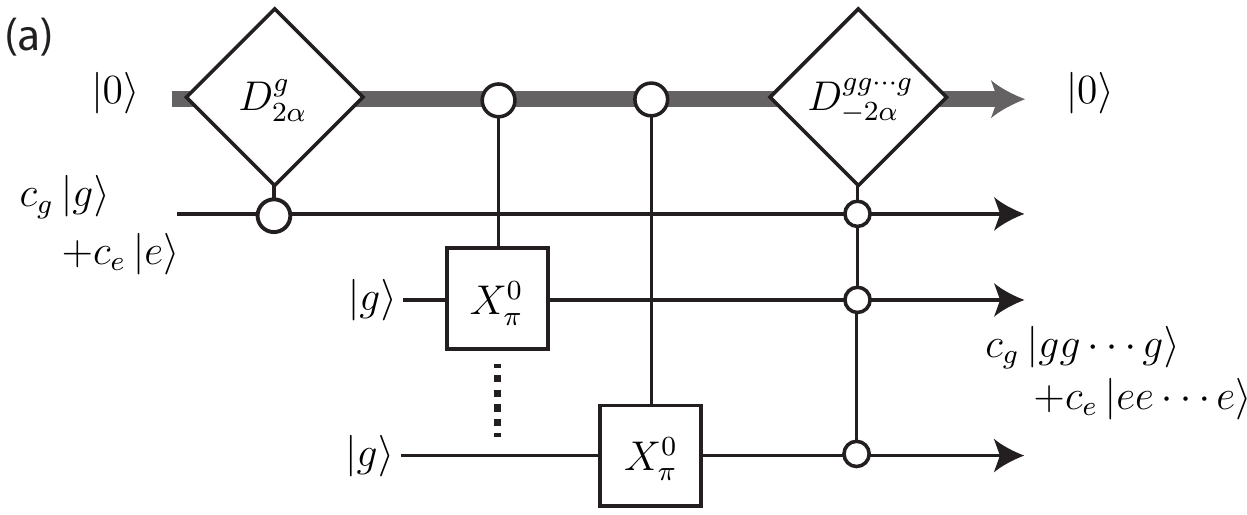}}
\put(0,0){\includegraphics[width=\columnwidth]{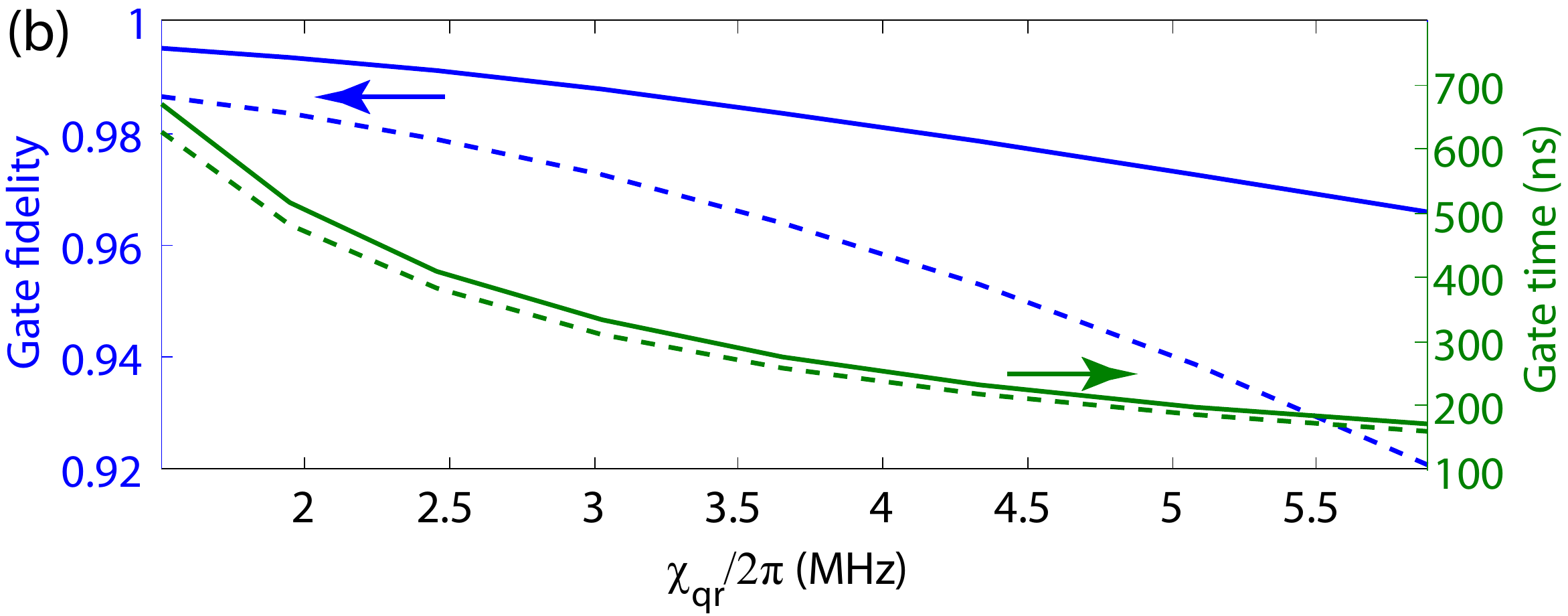}}
\end{picture}
\caption{\label{fig:GHZ} (a) The qcMAP gate can be used to map a single qubit state $c_g\ket{g}+c_e\ket{e}$ to a GHZ-type state $\ket{\text{GHZ}}=c_g\ket{gg\cdots g}+c_e\ket{ee\cdots e}$ for an arbitrary number of qubits; The conditional rotations of qubits can be done in parallel and therefore the total preparation time does not increase with the number of qubits $n_q$ (it actually slightly decreases with $n_q$ since the conditional displacement $D^{gg\cdots g}_{-2\alpha}$ can be performed faster). (b) Gate fidelity (blue) and time (green) as a function of the dispersive coupling strength for $3$ (solid lines) and $5$ qubits (dashed lines); we take the same dispersive shifts $\chi_{qr}/2\pi=3$ MHz for all qubits (not a necessary assumption) and $|\alpha|^2=3.5$. Like in Fig.~\ref{fig:SwapGate}, the simulation does not include decoherence but takes into account the cavity self-Kerr. For larger $n_q$, the cavity self-Kerr increases which leads to a drop in gate fidelity, particularly for high dispersive coupling strengths. We obtain fidelities in excess of $99\%$ (resp. $98\%$) for $n_q=3$ (resp. $n_q=5$) with a gate time of $400$~ns.}
\end{figure}

We have shown that the qcMAP gate  generates highly non-classical cavity field states, making it a promising tool to store multi-qubit states in the cavity~\cite{gottesman-et-al-01,vitali-et-al-98,zippilli-et-al-03}. An extension of the qcMAP gate uses the cavity as a bus to perform \emph{multi}-qubit gates. As shown in Fig.~\ref{fig:GHZ}(a), starting from state $c_g\ket{g}+c_e\ket{e}$ for one qubit, we use the qcMAP gate to map this state to a multi-qubit entangled state $c_g\ket{gg\cdots g}+c_e\ket{ee\cdots e}$. A first conditional displacement $D_{2\alpha}^{g}$ prepares $c_g\ket{2\alpha,gg\cdots g}+c_e\ket{0,eg\cdots g}$. The time for this operation is $\approx\pi/\chi_1$. Applying, in parallel, a conditional $\pi$-pulse $X_{\pi}^0$ on $n_q-1$ qubits, we prepare an $(n_q+1)$-body entangled state $c_g\ket{2\alpha,gg\cdots g}+c_e\ket{0,ee\cdots e}$. The time for this operation is fixed by the minimum dispersive coupling strength. Next, we perform a conditional displacement $D_{-2\alpha}^{gg\cdots g}$ disentangling the cavity from the qubits which are left in $c_g\ket{gg\cdots g}+c_e\ket{ee\cdots e}$, while the cavity is in vacuum. This conditional displacement can be performed in a very short time $\approx\pi/(\chi_1+\cdots+\chi_{n_q})$, which decreases with the number of qubits. Such an operation can be compared to the joint readout of qubits in the strong dispersive regime~\cite{filipp-et-al:PRL2009,chow-et-al:PRA2010} where, by driving the cavity at a frequency corresponding to a particular joint state of qubits, one can measure its population with a high fidelity.

In Fig.~\ref{fig:GHZ}(b), we plot the gate time and fidelity as a function of the dispersive coupling $\chi_{qr}$. A limiting effect on the fidelity is the cavity self-Kerr which increases additively with the number of qubits. Despite this effect, for $\chi_{qr}/2\pi=3~$MHz, we prepare a $5$-qubit GHZ state with a $\approx 97.5\%$ fidelity in $300$~ns. Furthermore, this gate can be performed between any subset of qubits coupled to the bus and does not require any qubit tunability or employment of higher excited states.

In conclusion, we have introduced the qcMAP gate which maps a qubit state to a superposition of two coherent states in a cavity.
The qcMAP gate is then used to prepare 2, 3 and 4 component SQOCS, as well as a non-local mesoscopic field state superposition in two cavity modes. Using this gate, the resonator could be used as a quantum ``disc drive'' to store \emph{multi}-qubit states in a multi-component SQOCS. A SQOCS of maximum photon number $\bar n$, for which the maximum non-orthogonality of two coherent components is $e^{-4m}$, could store a register of $\approx\log_2(\bar n/m)$ qubits. The effective decay rate of such a state would be $\bar n \kappa$ where $\kappa$ is the decay rate of one photon.
Using the qcMAP gate, the cavity can also be used as a bus to perform a multi-qubit gate, preparing in particular, GHZ states. Finally, any multi-qubit gate can be performed by concatenating such qcMAP gates.

This work was partially supported by the French ``Agence Nationale de la Recherche'' under the project EPOQ2 number ANR-09-JCJC-0070 and
the Army Research Office (ARO) under the project number ARO - W911NF-09-1-0514. ZL acknowledges support from the Fondation Sciences Math\'ematiques de Paris.


\begin{widetext}
\vspace{14cm}
\begin{center}
\large\textbf{Supplemental material for: deterministic protocol for mapping a qubit to coherent state superpositions in a cavity}
\end{center}
\end{widetext}

In this supplemental material, we describe in detail the $D_\alpha^g$ operation and provide the full sequence of steps that prepares the two, three and four component superposition of quasi-orthogonal coherent states with performances announced in the letter. Finally, a simple computation shows the first order effect of the cavity self-Kerr on a coherent state.

While the conditional qubit rotation $X_\theta^0$ is performed through long enough pulses ensuring a selective addressing of spectral lines (see letter), the conditional cavity displacement $D_\alpha^g$ is composed of two short unconditional displacements separated by a waiting time. This reduces the $D_\alpha^g$ operation time from $\approx 30/\chi_{qr}$ to $\approx\pi/\chi_{qr}$. We consider the rotating frame of the Hamiltonian $\omega_r a^\dag a+\omega_q b^\dag b-\frac{\chi_{qq}}{2}(b^\dag b)^2$. We perform a first unconditional displacement  $D_{\beta}$ of the cavity through a very short pulse that displaces the cavity regardless of the qubit state.  We wait for time $T_{\text{wait}}$, and apply a second unconditional displacement $D_{-\beta e^{i\chi_{qr}T_{wait}}}$. Neglecting the cavity self-Kerr, this sequence of operations leads to the following  unitary evolution:
\begin{align*}
\mathcal{U} &=D_{-\beta e^{i\chi_{qr}T_{wait}}}e^{i\chi_{qr}T_{\text{wait}}a^\dag a b^\dag b}D_{\beta}\\
&=e^{-i|\beta|^2\sin(\chi_{qr}T_{wait})}\ket{g}\bra{g}\otimes D_{\beta-\beta e^{i\chi_{qr}T_{wait}}}\\
& \qquad \qquad\qquad\qquad\qquad+\ket{e}\bra{e}\otimes e^{-i\chi_{qr}T_{\text{wait}}a^\dag a}.
\end{align*}
Taking $\alpha=\beta-\beta e^{i\chi_{qr}T_{wait}}$, we have
\begin{align*}
\mathcal{U}\ket{g,0}&=e^{-i|\beta|^2\sin(\chi_{qr}T_{wait})}\ket{g,\alpha},\\
\mathcal{U}\ket{e,0}&=\ket{e,0}.
\end{align*}
Up to a phase term of $e^{-i|\beta|^2\sin(\chi_{qr}T_{wait})}$ that we can take into account in future qubit pulses, this is precisely the conditional displacement. In particular, taking $T_{\text{wait}}=\pi/\chi_{qr}$, we have a conditional displacement of $\alpha=2\beta$ in a time of $\pi/\chi_{qr}$.

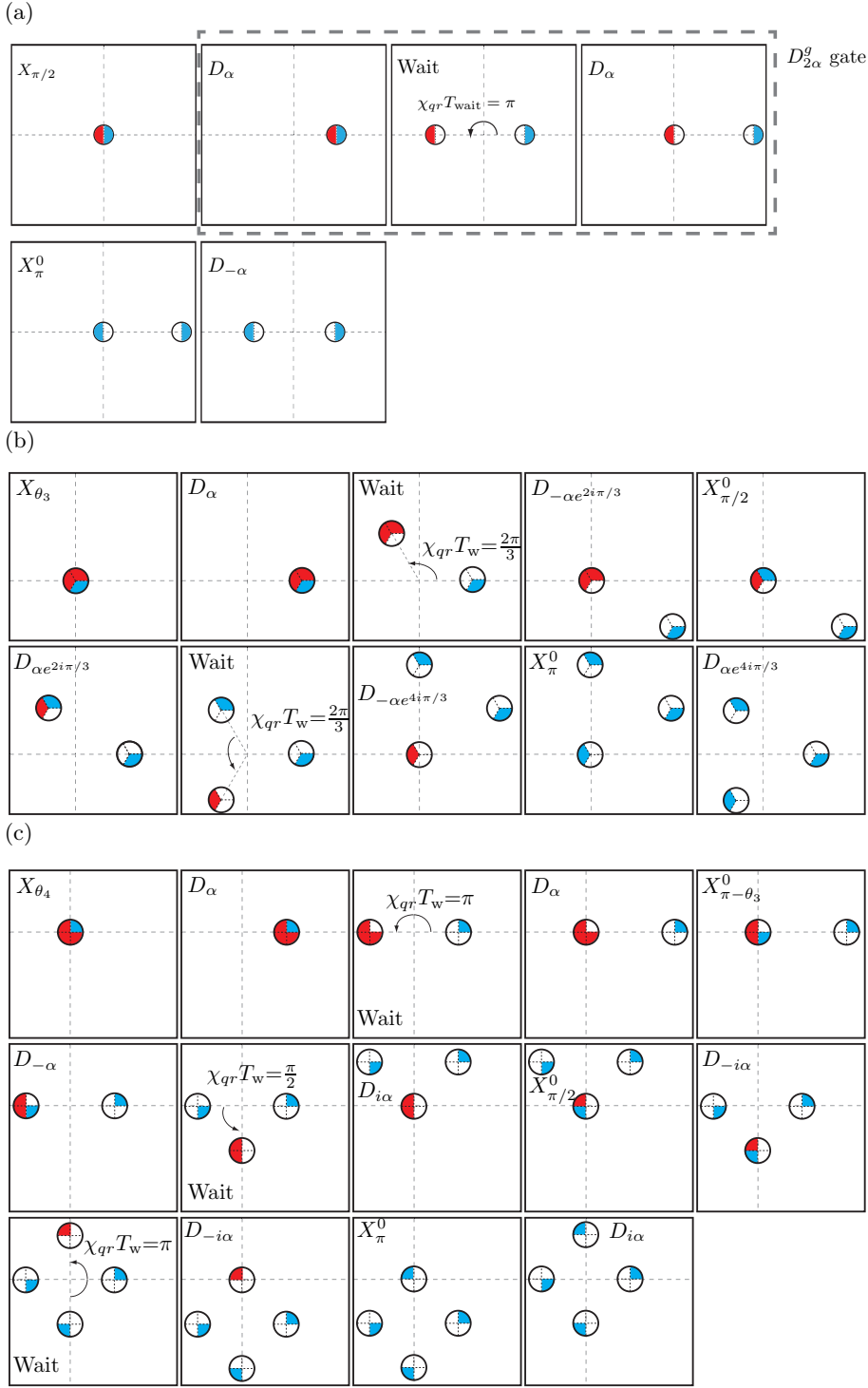
\begin{figure*}
\setlength{\unitlength}{1mm}
\begin{center}
\begin{picture}(220,200)
\put(20,192){(a)}
\put(20,135){\includegraphics[width=1.4\columnwidth]{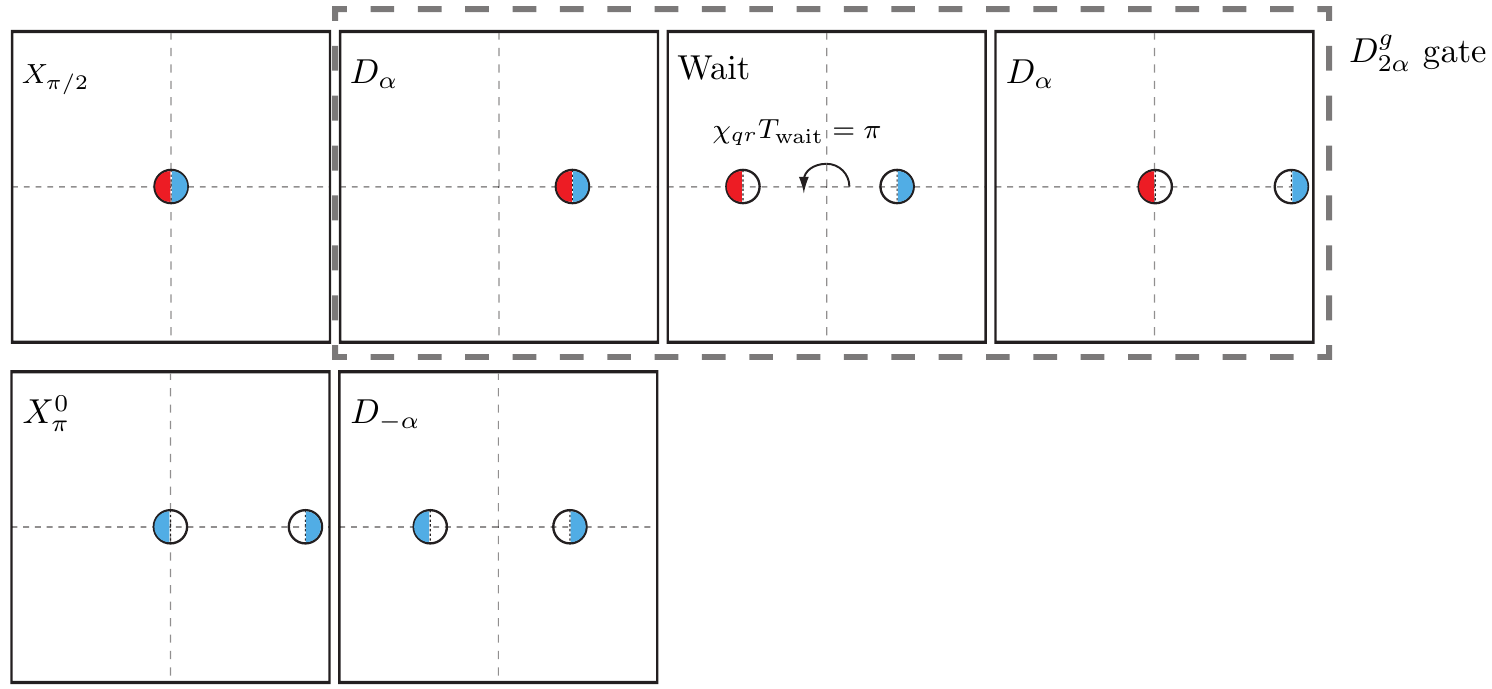}}
\put(20,132){(b)}
\put(20,80){\includegraphics[width=1.4\columnwidth]{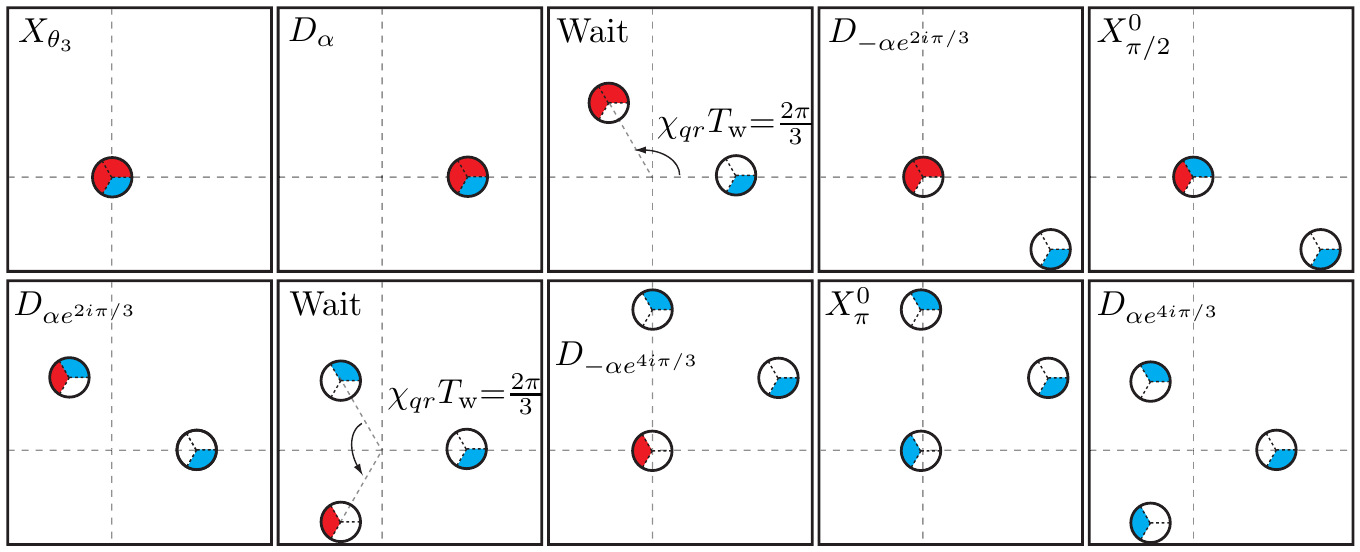}}
\put(20,77){(c)}
\put(20,0){\includegraphics[width=1.4\columnwidth]{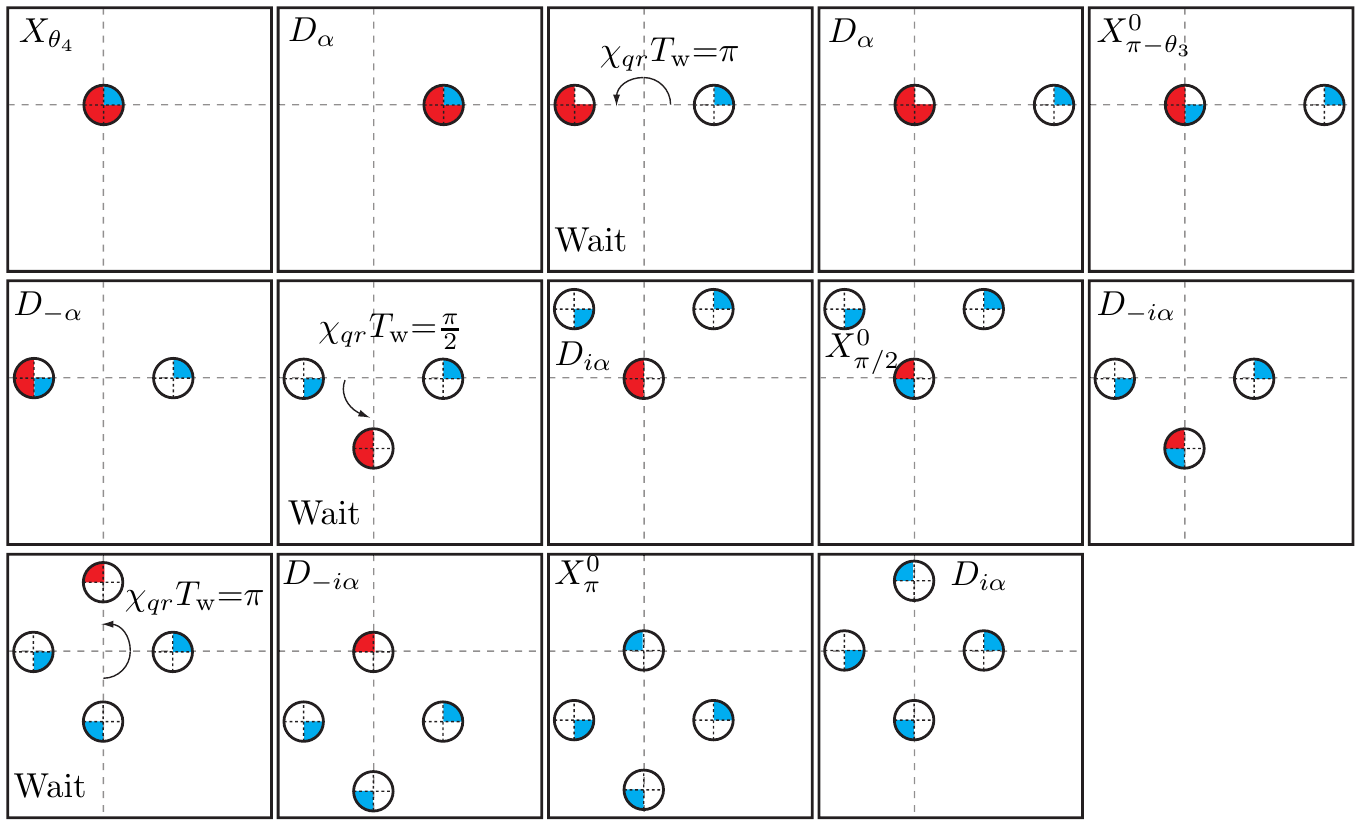}}
\end{picture}
\end{center}
\caption{\label{fig:DetailedSequences} Detailed sequence to prepare a superposition of  $2$ (a), $3$ (b) and $4$ (c)  quasi-orthogonal coherent states. Each frame is the Fresnel diagram of the field in the resonator. The two dotted lines represent two orthogonal quadratures, and intersect at $0$. The frames are ordered from left to right and top to bottom. A circle of center $\alpha$ in the diagram refers to a coherent state of amplitude $\alpha$. The fraction of the circle colored in blue [resp: red] corresponds to the population of the qubit which is in the ground state [resp: excited state]. Eg. frame 3 in (a) corresponds to state $\tfrac{1}{\sqrt{2}}(\ket{g,\alpha}+\ket{e,-\alpha})$. In particular we represent $\tfrac{1}{\sqrt{2}}\ket{g,\alpha}$ with the right circle ($+\alpha$), with a qubit in $\ket{g}$ (blue color) and a $50\%$ population (half full). Fast (here considered instantaneous) displacements $D_\gamma$ transform any coherent state $\ket{\alpha}$ to $\ket{\alpha+\gamma}$ regardless of the qubit state. The Fresnel diagram is in a rotating frame which leaves states of the form $\ket{g,\alpha}$ unchanged, while $\ket{\psi(0)}=\ket{e,\alpha}$ evolves as $\ket{\psi(t)}=\ket{e,\alpha e^{i\chi_{qr}t}}$. A selective pulse $X_\theta^0$ rotates the qubit state when the resonator is in the zero photon state $\ket{0}$. Graphically, this corresponds to changing a fraction of the color of a circle centered at $0$. In (b) and (c), $\theta_3=2\arccos(1/\sqrt 3)$ and in (c) $\theta_4=2\arccos(1/2)$. A symbol in each frame $n$ gives the operation performed to go from frame $n-1$ to $n$.}
\end{figure*}

In Figure~\ref{fig:DetailedSequences}, we provide the complete sequence of operations which generate superpositions of two, three and four quasi-orthogonal coherent components.

Let us finish by a simple computation showing the first order effects of the cavity self-Kerr. Considering a short time $\tau$ such that $\epsilon=\chi_{rr}\tau/2\ll 1$, we can show that the first order contribution of the cavity self-Kerr is simply an extra deterministic phase accumulation of the cavity's coherent states that we can take into account in future cavity displacements and qubit rotations. Indeed, the distortion of the coherent states happens only as a second order term with respect to $\epsilon$. Consider a coherent state $\ket{\alpha}$ of average photon number $\bar n=|\alpha|^2$. We define $\ket{\psi_\epsilon}=e^{i\epsilon (a^\dag a)^2}\ket{\alpha}$ and we search for a coherent state of amplitude $\alpha_\epsilon$ and global phase $\phi_\epsilon$: $e^{i\phi_\epsilon}\ket{\alpha_\epsilon}$, which is close to $\ket{\psi_\epsilon}$ for small $\epsilon$. We have $\bra{\psi_\epsilon}a\ket{\psi_\epsilon}=\alpha e^{i\epsilon}e^{\bar n(e^{2i\epsilon}-1)}=\bra{\alpha_\epsilon}a\ket{\alpha_\epsilon}$ for $\alpha_\epsilon=\alpha e^{i\epsilon}e^{\bar n(e^{2i\epsilon}-1)}=\alpha e^{i\epsilon(2\bar n+1)}+O(\epsilon^2)$. In order to find $\phi_\epsilon$, we compute
\begin{eqnarray*}
  e^{-i\phi_\epsilon}\scprod{\alpha_\epsilon}{\psi_\epsilon}&=&e^{-i\phi_\epsilon}\bra{\alpha}e^{i\epsilon(-(2\bar n+1)a^\dag a+(a^\dag a)^2)}\ket{\alpha}+O(\epsilon^2)\\
  &=&e^{-i\phi_\epsilon}(1-i\epsilon\bar n^2)+O(\epsilon^2)
\end{eqnarray*}
Taking $\phi_\epsilon=-\epsilon\bar n^2$, we get $e^{-i\phi_\epsilon}\scprod{\alpha_\epsilon}{\psi_\epsilon}=1+O(\epsilon^2)$.
Therefore, as a first order approximation for the effect of the cavity self-Kerr, we have
$$
e^{i\chi_{rr}\tau (a^\dag a)^2/2}\ket{\alpha}\sim e^{-i\chi_{rr}\tau  \bar n^2/2}\ket{e^{i\chi_{rr}\tau(\bar n+\frac{1}{2})}\alpha}.
$$
In the simulations of this letter, we took into account this extra coherent state rotation for the following displacements. The overall phases were corrected by adequately choosing the following qubit pulse phases.

\end{document}